\documentclass{new_tlp}
\usepackage{mathptmx}

\usepackage{subfigure}

\usepackage[most]{tcolorbox}
\usepackage{makeidx}
\usepackage[english]{babel}
\usepackage{amsmath}
\usepackage{amsfonts}
\usepackage{amssymb}
\usepackage{graphicx}
\usepackage{xspace}
\usepackage{url}
\usepackage{algorithm}
\usepackage{algpseudocode}
\usepackage{booktabs}
\usepackage{multirow}
\usepackage[labelsep=quad,indention=10pt]{subfig}
\usepackage{bbding}
\usepackage{todonotes}
\usepackage{adjustbox}
\usepackage{tikz}
\usepackage{scalefnt}
\usepackage{hyphenat}
\usepackage{pgfplots}
\usepackage[T1]{fontenc}

\usetikzlibrary{plotmarks}
\pgfplotsset{
	filter discard warning=false 
	, legend cell align=left
	, minor grid style={loosely dotted, lightgray}
	, major grid style={loosely dashed, lightgray}
}

\newcommand{\dlv}{\textsc{dlv}\xspace}
\newcommand{\dlva}{\textsc{dlv}}
\newcommand{\idlv}{{{\small $\cal I$}-}\dlv}
\newcommand{\idlva}{{{\small $\cal I$}-}\dlva}
\newcommand{\idlvs}{\idlva+{\footnotesize \ensuremath{ \mathcal{MS}}}\xspace}

\newcommand{\cmodels}{\textsc{cmodels}\xspace}
\newcommand{\wasp}{\textsc{wasp}\xspace}
\newcommand{\waspa}{\textsc{wasp}}
\newcommand{\waspmod}{\waspa$^*$}
\newcommand{\clasp}{\textsc{clasp}\xspace}
\newcommand{\claspa}{\textsc{clasp}}
\newcommand{\claspmod}{\claspa$^*$}

\newcommand{\measp}{\textsc{measp}\xspace}
\newcommand{\gringo}{\textsc{gringo}\xspace}
\newcommand{\gringoa}{\textsc{gringo}}
\newcommand{\clingo}{\textsc{clingo}\xspace}

\newcommand{\lparse}{\textsc{lparse}\xspace}

\newcommand\mydots{\hbox to .8em{.\hss.\hss.}}
\newcommand\smaller[2][0.85]{{\scalefont{#1}#2}}
\newcommand{\dlvdue}{\textsc{dlv\smaller[0.8]{2}}\xspace}

\newcommand{\preproc}{{\small\textsc{Preprocessor}}\xspace}
\newcommand{\stats}{{\small\textsc{Grounding analyser}}\xspace}
\newcommand{\selector}{{\small\textsc{Solver Selector}}\xspace}

\newcommand{\naf}{\ensuremath{not\ }}

\newcommand{\cit}[1]{~\cite{#1}}
\newcommand{\re}[1]{~\ref{#1}}

\newcommand{\commentsymbolright}{/*}
\newcommand{\commentsymbolleft}{*/}
\algrenewcommand\algorithmiccomment[1]{\hfill \commentsymbolright{} #1 \commentsymbolleft{}}


\begin{document}
\bibliographystyle{acmtrans}

\title[Efficiently Coupling the I-DLV Grounder with ASP Solvers]
    {Efficiently Coupling the I-DLV Grounder \\ with ASP Solvers}

\author[Francesco Calimeri et al.]
    {FRANCESCO CALIMERI, DAVIDE FUSC\`{A}, SIMONA PERRI, JESSICA ZANGARI\\
    	Department of Mathematics and Computer Science, University of Calabria, Italy\\
    	\email{$\{$calimeri,fusca,perri,zangari$\}$@mat.unical.it}\\    	
    	\and CARMINE DODARO\\
		Department of Informatics, Bioengineering, Robotics and Systems Engineering, University of Genova, Italy \\
		\email{dodaro@dibris.unige.it}
    }

\maketitle

\begin{abstract}


We present \idlvs, a new Answer Set Programming (ASP) system that integrates an efficient grounder, namely \idlv, with an automatic selector that inductively chooses a solver: depending on some inherent features of the instantiation produced by \idlv, machine learning techniques guide the selection of the most appropriate solver.
The system participated in the latest ($7th$) ASP competition, winning the regular track, category $SP$ (i.e., one processor allowed).
Under consideration in Theory and Practice of Logic Programming (TPLP). 

\end{abstract}

\section{Introduction}\label{sec:intro}

Answer Set Programming (ASP)~\cite{DBLP:journals/cacm/BrewkaET11,DBLP:journals/ngc/GelfondL91} is a declarative programming paradigm proposed in the area of non-monotonic reasoning and logic programming.
The language of ASP is based on rules, allowing (in general) for both disjunction in rule heads and nonmonotonic negation in the bodies; such programs are interpreted according to the answer set semantics~\cite{DBLP:conf/iclp/GelfondL88,DBLP:journals/ngc/GelfondL91}.
Throughout the years the availability of reliable, high-performance implementations~\cite{DBLP:journals/ai/CalimeriGMR16,DBLP:conf/aaai/GebserMR16} made ASP a powerful tool for developing advanced applications in many research areas, ranging from artificial intelligence to databases and bioinformatics, as well as in industrial contexts~\cite{DBLP:journals/ai/CalimeriGMR16,DBLP:conf/rweb/LeoneR15,DBLP:conf/padl/NogueiraBGWB01,DBLP:conf/rr/DodaroLNR15,DBLP:journals/tplp/RiccaGAMLIL12,tiihonen2003practical,DBLP:conf/lpnmr/GrassoILR09,DBLP:journals/tplp/DodaroGLMRS16}.

The ``traditional'' approach to the evaluation of ASP programs relies on a grounding module (\emph{grounder}) that generates a propositional theory semantically equivalent to the input program coupled with a subsequent module (\emph{solver}) that uses propositional techniques for generating answer sets.
There have been other attempts deviating from this customary approach~\cite{DBLP:journals/fuin/PaluDPR09,DBLP:journals/tplp/LefevreBSG17,DBLP:conf/jelia/Dao-TranEFWW12,DBLP:conf/lpnmr/Weinzierl17}; nonetheless, the majority of the current solutions relies on the canonical ``ground \& solve'' strategy, including \dlv~\cite{DBLP:conf/lpnmr/AlvianoCDFLPRVZ17,DBLP:journals/tocl/LeonePFEGPS06} and \clingo~\cite{DBLP:conf/lpnmr/GebserKK0S15,DBLP:conf/lpnmr/GebserKKS11}.

\medskip

While a significant amount of solvers has been released in the latest years, as confirmed by the most recent editions of the ASP competitions~\cite{DBLP:journals/ai/CalimeriGMR16,DBLP:journals/jair/GebserMR17,DBLP:conf/lpnmr/GebserMR17}, until recently, the  alternatives for grounding basically consisted of \gringo~\cite{DBLP:conf/lpnmr/GebserKKS11} and the grounding module of the \dlv system~\cite{DBLP:journals/tocl/LeonePFEGPS06,DBLP:conf/birthday/FaberLP12}; furthermore, the \dlv grounder was conceived as part of a monolithic system supporting a specific input language, thus making interfacing with other solvers difficult: this led to a wide adoption of \gringo as a base for a number of ASP solutions.
Recently, \idlv~\cite{DBLP:journals/ia/CalimeriFPZ17} has been released as a new ASP grounder featuring high flexibility and customizability, and a lightweight modular design for easing the incorporation of optimization techniques and updates;
\idlv can be easily combined with different solvers, since it supports the ASP-Core-2 standard language~\cite{calimeri2012asp} as well as the de-facto standard numerical format for ground programs~\cite{lparse-manual}; hence, it constitutes a valid alternative to \gringo.

The \idlv performance has already been assessed, proving that it is competitive  both as grounder and as deductive database system~\cite{DBLP:conf/padl/CalimeriFPZ18,DBLP:journals/ia/CalimeriFPZ17}. However, the impact on ASP solvers of the instantiation it produces  is still largely unexplored. We started a first experimental analysis in this respect by coupling \idlv with the two state of the art solvers \clasp~\cite{DBLP:conf/lpnmr/GebserKK0S15} and \wasp~\cite{DBLP:conf/lpnmr/AlvianoDLR15}.
Furthermore, given that, due to possibly different heuristics, optimization techniques, data structures, or input simplifications, such solvers outperform each other depending on the ground program produced by \idlv given the input at hand~\cite{DBLP:journals/jair/GebserMR17},
we explored possible ways of developing an efficient ASP system that was able to take advantage from peculiarities of both.
To this aim, we designed an ad hoc machinery for dynamically selecting the solver to be coupled with \mbox{\idlv} on an instance basis.
The resulting system features a dynamic selector that relies on specific metrics that somehow ``describe'' the form of the \idlv instantiations.
In order to build the selector, we first introduced a specific set of features for defining the metrics, and then we carried out an experimental activity for actually computing them over all benchmarks submitted to the $6th$ ASP Competition~\cite{DBLP:journals/jair/GebserMR17} as grounded by \idlv.
We then performed an extensive experimental evaluation in order to first select the most promising features, and then choose the best performing selection algorithm among four different approaches, namely: Random Forest (RF), Neural Network (NN), Support Vector Machines (SVM) and AutoWEKA.
The result is the new ASP system \idlvs, which participated in the latest ($7th$) ASP competition~\cite{DBLP:conf/lpnmr/GebserMR17} winning in the regular track, category $SP$ (i.e., one processor allowed), and is the subject of the present work, where its performance is also assessed by means of proper experimental analyses.


\medskip

In the remainder of the paper we provide the reader with a brief overview of Answer Set Programming evaluation (Section~\ref{sec:evalasp}) before presenting some insight about \idlv (Section~\ref{sec:idlv}); we then describe the design and implementation of \idlvs
(Section~\ref{sec:selector}), and the results of a proper experimental evaluation
(Section~\ref{sec:experiments}).
We discuss related work in Section~\ref{sec:related} before illustrating ongoing works and drawing our conclusions (Section~\ref{sec:conclusions-ongoing}).


\section{Evaluation of ASP programs}\label{sec:evalasp}
We briefly recall that the core of the ASP language consists in disjunctive Datalog with nonmonotonic negation under the stable model semantics. Nevertheless, over the years a significant amount of work has been carried out by the scientific community in order to enrich the basic language, and several extensions have been studied and proposed. Recently, the community agreed on a standard input language for ASP systems: ASP-Core-2~\cite{calimeri2012asp}, the official language of the ASP Competition series~\cite{DBLP:journals/ai/CalimeriGMR16,DBLP:conf/aaai/GebserMR16,DBLP:conf/lpnmr/GebserMR17}. For a thorough reference on ASP syntax and semantics, we refer the reader to~\cite{calimeri2012asp}.

The state-of-the-art strategy for the evaluation of ASP programs\cit{DBLP:journals/aim/KaufmannLPS16} relies on two computational phases. In particular, this approach mimics the definition of the semantics by relying on a grounding module (\emph{grounder} or \emph{instantiator}) coupled with a subsequent module (\emph{solver}) that applies proper propositional techniques for generating its answer sets.
These phases are usually referred to as {\em instantiation} or {\em grounding}, and {\em solving} or {\em answer sets search}, respectively. The two phases are, in general, computationally expensive~\cite{DBLP:journals/csur/DantsinEGV01}, and efficient ASP computation depends on proper optimization of both.


Given a (non-ground) ASP program $P$, the grounding phase consists in producing a propositional theory $G_{P}$ semantically equivalent to $P$, i.e. such that $G_P$ does not contain any variable but has the same answer sets as $P$.
State-of-the-art grounders are based on semi-na\"{i}ve database evaluation techniques~\cite{DBLP:books/cs/Ullman88}, for avoiding duplicate work, and build predicate domains dynamically.
Basically, they adopt an iterative bottom-up process that successively expands the set of variable-free terms constructible from the non-ground program.

The core of the process is the rule instantiation step: given a rule $r$ and a set of ground (i.e. without variables) atoms $S$, which collects all the ground atoms appearing in the potential extensions of the predicates in $P$, it generates the ground instances of $r$.
By potential extension of a predicate $p$ we mean here the set of ground instances of $p$ that are either already known to be true in all possible answer sets, or that can possibly be true in some answer set; hence, these constitute the basis for generating the ``relevant'' ground rules that will undergo the solving phase.
In more detail, such task is performed by iterating on the positive body literals and matching them with ground atoms in $S$ so that possible substitutions for their variables are generated.
Indeed, because of the safety condition, it is enough to have a substitution for the variables occurring in the positive literals in order to generate a completely ground instance of $r$.
Notably, the set $S$ of predicate extensions is built dynamically: the rules are evaluated according to a specific order ensuring that the complete extension of all predicates needed for instantiating the current rule is already available.
The collection of the ground rules generated from all input rules in $P$ composes the ground program $G_P$.
Besides this basic schema, efficient grounders employ smart optimizations techniques, geared towards efficiently producing a ground program that is considerably smaller than the full instantiation, still preserving the semantics.


The propositional program produced by the grounder is evaluated by the solver, whose role is to produce answer sets.
Modern ASP solvers implement the backtracking algorithm CDCL, which is based on the pattern \textit{choose}-\textit{propagate}-\textit{learn}. Intuitively, the idea is to build an answer set step-by-step by starting from an empty set of literals $\mathcal{I}$. Then, the algorithm heuristically chooses a literal to be added in $\mathcal{I}$, and the deterministic consequences of this choice are propagated, i.e. new literals are possibly added to $\mathcal{I}$.
The propagation may lead to a conflict, i.e. a literal and its negation are both in $\mathcal{I}$.
In this case, the conflict is analyzed and a new constraint is added to the propositional program (learning).
The conflict is then repaired, i.e. $\mathcal{I}$ is restored in a consistent state, choices leading to the conflict are retracted, and a new literal is heuristically selected to be added in $\mathcal{I}$.
The algorithm then iterates until $\mathcal{I}$ contains either $p$ or $\naf p$, for each atom $p$ appearing in the input program, i.e. an answer set is produced, or the incoherence of the propositional program is proved, i.e. no answer sets are admitted.

\section{The \idlv Grounder}\label{sec:idlv}


The recently released system \idlv~\cite{DBLP:journals/ia/CalimeriFPZ17} is a brand new ASP grounder and full-fledged deductive database engine. Besides being a stand-alone system, it has been integrated as the grounder module of the new version of the popular ASP system \dlv, namely \dlvdue~\cite{DBLP:conf/lpnmr/AlvianoCDFLPRVZ17}.  Among the most widely used ASP systems, \dlv~\cite{DBLP:journals/tocl/LeonePFEGPS06} has been one of the first solid and reliable; its project dates back a few years after the first definition of answer set semantics~\cite{DBLP:conf/iclp/GelfondL88,DBLP:journals/ngc/GelfondL91}, and encompassed the development and the continuous enhancements of the system; \dlv is widely used in academy, and in many relevant industrial applications, significantly contributing in spreading the use of ASP in real-world scenarios~\cite{DBLP:journals/aim/ErdemGL16}.

In this section, after a brief overview on \idlv and its main features, we discuss the result of an experimental activity carried out in order to assess the impact of \idlv on modern ASP solvers.

\subsection{Overview}

\idlv has been redesigned and re-engineered from scratch in order to natively supports ASP-Core-2~\cite{calimeri2012asp} and the latest technologies.
Its grounding process is based on the basic instantiation schema briefly described above (see Section~\ref{sec:evalasp}), that is conveniently enhanced by a number of optimization techniques explicitly designed in the context of ASP instantiation.
In particular, \idlv employs an improved version of already known techniques having a high impact on the overall instantiation process, such as {\em body-reordering} and \emph{indexing} strategies, along with several additional fine-tuning optimizations acting to different extents on the instantiation process, with the general common aim of reducing the search space and improving overall performance.

Furthermore, as the same computational problem can be encoded by means of many different ASP programs which are semantically equivalent, it is well-known that real ASP systems may perform very differently when evaluating each one of them.
This behavior is due, in part, to specific aspects, that strictly depend on the ASP system  employed, and, in part, to general ``intrinsic'' aspects, depending on the program at hand which could feature some characteristics that can make computation easier or harder.
Thus, often, to have satisfying performance, expert knowledge is required in order to select the best encoding, and this, in a certain sense,  conflicts with the declarative nature of ASP that, ideally, should free the users from the burden of the computational aspects.
For this reason, \idlv is endowed with proper means aiming at making performance less encoding-dependent; intuitively, this is crucial for fostering and easing the usage of ASP in practice.
One of the more effective techniques, which has been recently introduced into \idlv,
is the {\em decomposition rewriting}~\cite{DBLP:conf/padl/CalimeriFPZ18}.
The rationale behind this technique comes from the consideration that when programs contain rules featuring long bodies, ASP systems performance might benefit from a careful split of such rules into multiple, smaller ones.
However, while in some cases such decomposition is convenient, in other cases keeping the original rule is preferable; hence, \idlv relies on smart rewriting strategies based on \textit{tree-decomposition} algorithms.
Basically, each input rule is analyzed before the evaluation, and the system estimates whether it is convenient to decompose it into an equivalent set of smaller rules, or not; if more than one decomposition is possible, the most promising is selected.
The method is general and defined so that all choices are made according to proper criteria and heuristics that can be customized: it can be tailored to different phases of the ASP computation, and it is not tied to a specific system.
The \idlv implementation features ad-hoc heuristics and criteria relying on data and statistics dynamically computed during the computation with the aim of optimizing the grounding performances.
It is worth noting that, among all \idlv optimizations, some have a significant impact only on instantiation times, while others may impact also on solving times, as they could lead to changes in size and structure of the produced ground program; decomposition rewriting is counted among the latter.

\idlv comes with a general-purpose default configuration for its optimizations (for a more comprehensive list and a more thorough discussion, we refer the reader to~\cite{DBLP:journals/ia/CalimeriFPZ17}); however, it also provides the user with means for a fine-grained control over the whole computational process.
Indeed, the long-lasting experience over the former \dlv grounder~\cite{DBLP:conf/birthday/FaberLP12} proved that a monolithic set of optimizations, most of which were activated or deactivated at the same time, does not pay in general; on the contrary, the possibility to independently set them allows to ad hoc optimize the computation and also tailor the produced instantiation to different extents.
The novel possibility of \emph{annotating ASP code} with external directives to the grounder is a further bold move in this direction.
In \idlv, annotations allow to give explicit directions on the internal computational process.
Currently, supported annotations belong to the following categories: \textit{grounding annotations} allowing for a fine-grained customization on the grounding process, and \textit{solving annotations} that have been integrated into \dlvdue, and are geared towards the customization of heuristics and extension of solving capabilities~\cite{DBLP:conf/lpnmr/AlvianoCDFLPRVZ17}.

As additional features, \idlv has been endowed with means to ease the interoperability with external sources of knowledge.
In particular, \idlv supports calls to Python functions via {\em external atoms} that permit to combine imperative-oriented computational aspects with ASP, and connections with relational and graph databases via explicit {\em directives} for importing/exporting data.

The system is envisioned as core part of a larger project comprising the extension of \idlv towards mechanisms for interoperability with other formalisms and tools: the intent is to foster the usage of ASP, and in general, of logic programming in real-world and complex applications.
Moreover, \idlv is an open-source project: its source and binaries are available from the official repository~\cite{idlv-web}.
For a comprehensive list of customizations and options, along with further details, we refer the reader to~\cite{DBLP:journals/ia/CalimeriFPZ17} and to the online documentation~\cite{idlv-web}.

\subsection{Impact of \idlv on ASP solvers}\label{idlv-impact}
In order to analyze the impact of the instantiation produced by \idlv on the solving phase, we carried out an ad-hoc experimental activity comparing performance of \idlv combined with the two mainstream solvers \clasp and \wasp.
Furthermore, we tested also the same solvers when coupled with the well-established ASP grounder \gringo, in order to have a solid term of comparison.

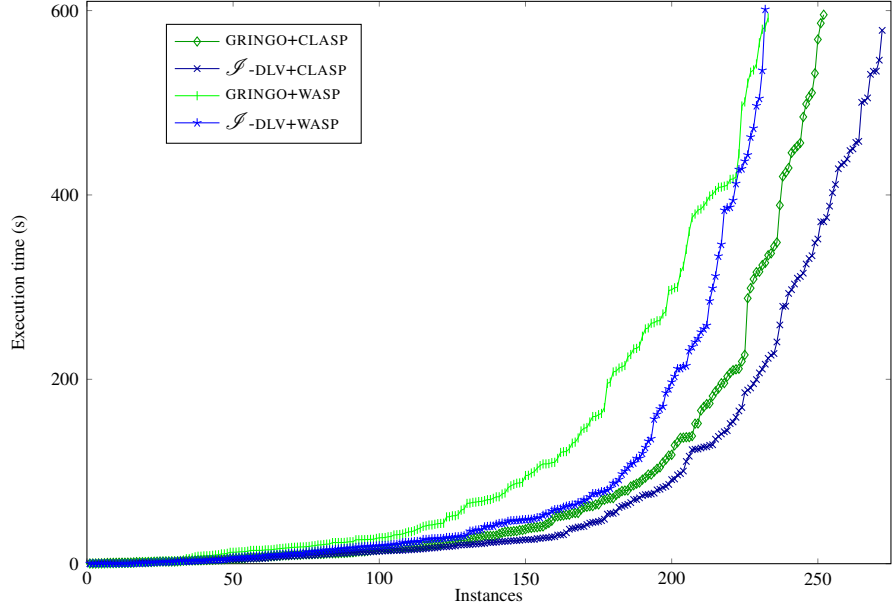
\begin{figure}[t!]\centering
\begin{tikzpicture}[scale=0.99]
	\pgfkeys{/pgf/number format/set thousands separator = {}}
    	\begin{axis}[
    	scale only axis
    	, font=\scriptsize
    	, x label style = {at={(axis description cs:0.5,0.04)}}
    	, y label style = {at={(axis description cs:0.03,0.5)}}
    	, xlabel={Instances}
    	, ylabel={Execution time (s)}
    	, width=0.8\textwidth
        , legend style={at={(0.22,0.96)},anchor=north}
    	, height=0.56\textwidth
    	, ymin=0, ymax=610
    	, xmin=0, xmax=275
    	, xtick={0,50,100,150,200,250}
    	, ytick={0,200,400,600}
    	, major tick length=2pt
    	]
    \addplot [mark size=1.75pt, color=green!60!black, mark=diamond] [unbounded coords=jump] table[col sep=semicolon, y index=1] {./sixth-comp-idlv-gringo-clasp-wasp-no-query.csv};
    	\addlegendentry{\gringoa+\clasp}
    \addplot [mark size=1.75pt, color=blue!60!black, mark=x] [unbounded coords=jump] table[col sep=semicolon, y index=3] {./sixth-comp-idlv-gringo-clasp-wasp-no-query.csv};
    	\addlegendentry{\idlva+\clasp}
    \addplot [mark size=1.75pt, color=green, mark=|] [unbounded coords=jump] table[col sep=semicolon, y
    index=2] {./sixth-comp-idlv-gringo-clasp-wasp-no-query.csv};
        \addlegendentry{\gringoa+\wasp}
    \addplot [mark size=1.75pt, color=blue, mark=star] [unbounded coords=jump] table[col sep=semicolon, y index=4] {./sixth-comp-idlv-gringo-clasp-wasp-no-query.csv};
        \addlegendentry{\idlva+\wasp}
        \end{axis}
	\end{tikzpicture}
\caption{Cactus plot representing the comparison of \gringoa+\clasp, \idlva+\clasp, \gringoa+\wasp and \idlva+\wasp on benchmarks from the $6th$ ASP Competition, where domains featuring queries have been removed: number of solved instances on the $x$-axis while running times (in seconds) are on the $y$-axis.\label{fig:solving}}
\end{figure}


All systems have been executed with their default configuration and the versions employed are the latest available at the time of executing the experiments:
\idlv {\ \small\texttt{2017-06-01}}, \gringo and \clasp as used by \clingo {\ \small\texttt{5.1.0}}, \wasp {\ \small\texttt{2017-05-04}}. The executables are available at~\cite{idlv-ms-web}.

We relied on the whole $6th$ ASP Competition suite~\cite{DBLP:journals/jair/GebserMR17}\footnote{The $7th$ ASP competition~\cite{DBLP:conf/lpnmr/GebserMR17} is already over at the time of writing, but was not held yet at the time we were designing \idlvs; furthermore, the benchmarks therein employed are not publicly available yet, at the time of writing.}
that features $28$ problems and $20$ different instances per each problem; problems are categorized into three types: $(i)$ \textit{decision} (encodings do not contain queries or weak constraints), $(ii)$ \textit{optimization} (weak constraints are present in the encodings), and $(iii)$ \textit{query} (encodings explicitly feature queries).

All experiments reported herein and in the following have been performed on a NUMA machine equipped with two \texttt{2.8 GHz AMD Opteron} \texttt{6320 CPUs} and \texttt{128 GiB} of main memory, running \texttt{Linux Ubuntu 14.04.4}; moreover, memory and time were limited to \texttt{15 GiB} and \texttt{600 seconds}, respectively, for each system per each single run.

\medskip

A first idea on the results can be given by Figure~\ref{fig:solving}, that shows a cactus plot illustrating the behaviour of the four considered combinations; here, domains featuring queries have been ignored, as they are not supported by \gringo: this should help at getting a more clear picture of the performance trends of the tested systems.
Number of solved instances are reported on the $x$-axis, while running times (in seconds) are on the $y$-axis.
It is evident that
%
\idlva+\clasp is the combination that performs better on the overall, solving the highest number of instances and resulting as the most efficient in the majority of the considered domains.
Furthermore, it is easy to see that, in general, both solvers benefit from the use of \idlv instead of \gringo.

A more detailed analysis can be done while looking at full data, reported in Table~\ref{tab:solver}, that shows average times and number of solved instances within the allotted time and memory; time outs and unsupported syntax are denoted by ``TO'' and ``US'', respectively.
%

\begin{table}[t]
  \centering
  \caption{Comparison of \gringoa+\clasp, \gringoa+\wasp, \idlva+\clasp and \idlva+\wasp on benchmarks from the $6th$ ASP Competition (20 instances per problem): number of solved instances and average running times (in seconds). Time outs and unsupported syntax issues are denoted by TO and US, respectively.
  \label{tab:solver}}
  \tabcolsep=0.040cm
  \begin{tabular}{lrrrrrrrr}
\hline\hline
     \multirow{2}{0.2\textwidth}{\textbf{Problem}} 	& \multicolumn{2}{c}{\textbf{\gringoa+\clasp}} 	& \multicolumn{2}{c}{\textbf{\gringoa+\wasp}} 	& \multicolumn{2}{c}{\textbf{\idlva+\clasp}} 	& \multicolumn{2}{c}{\textbf{\idlva+\wasp}} \\
     \cmidrule{2-3}     \cmidrule{4-5}      \cmidrule{6-7}      \cmidrule{8-9}
     	& \textbf{\#solved} 	& \textbf{time} 	& \textbf{\#solved} 	& \textbf{time} 	&\textbf{\#solved} 	& \textbf{time} 	&\textbf{\#solved} 	& \textbf{time} \\
\hline
    Abstract Dial. Frameworks 	& 20 	& 8.86 		& 12 	& 39.75 	& 20 	& 6.84 		& 11 	& 33.79 \\
    Combined Configuration 	& 9  	& 252.78 	& 1	& 1.17 		& 10 	& 177.19 	& 0  	& TO \\
    Complex Optimization 	& 17 	& 134.10 	& 5	& 199.83 	& 18 	& 159.52 	& 6  	& 152.15 \\
    Connected Still Life 	& 6  	& 245.75 	& 12	& 52.65 	& 6  	& 253.13 	& 12 	& 58.25 \\
    Consistent Query Ans. 	& 0  	& US 		& 0	& US 		& 20 	& 86.37 	& 18 	& 87.34 \\
    Crossing Minimization 	& 6  	& 64.32 	& 19	& 4.17 		& 7  	& 55.88 	& 19 	& 5.52 \\
    Graceful Graphs 		& 9 	& 70.68 	& 4	& 47.69 	& 9  	& 136.99 	& 6  	& 194.57 \\
    Graph Coloring 		& 15 	& 142.25 	& 8	& 134.27 	& 15 	& 169.46 	& 8  	& 133.42 \\
    Incremental Scheduling 	& 12 	& 88.65 	& 5	& 130.09 	& 14 	& 120.34 	& 6  	& 77.94 \\
    Knight Tour With Holes 	& 10 	& 15.23 	& 10	& 112.84 	& 11 	& 58.90 	& 10 	& 34.85 \\
    Labyrinth 			& 13 	& 108.26 	& 10	& 139.45 	& 12 	& 148.17 	& 10 	& 70.98 \\
    Maximal Clique 		& 0  	& TO 		& 8  	& 310.99 	& 0  	& TO 		& 9 	& 353.50 \\
    MaxSAT 			& 7  	& 42.71 	& 19 	& 100.71 	& 7  	& 38.29 	& 19 	& 90.53 \\
    Minimal Diagnosis 		& 20 	& 8.45 		& 20 	& 35.87 	& 20 	& 8.70 		& 20 	& 24.45 \\
    Nomistery 			& 7  	& 92.30 	& 8	& 58.91 	& 8  	& 140.89 	& 8 	& 36.73 \\
    Partner Units 		& 14 	& 35.46 	& 10	& 227.77 	& 14 	& 20.43 	& 5  	& 131.61 \\
    Permut. Pattern Matching 	& 11 	& 165.67 	& 20	& 176.26 	& 20 	& 16.46 	& 20 	& 22.31 \\
    Qualitat. Spatial Reasoning	& 18 	& 116.66 	& 16	& 194.69 	& 20 	& 125.38 	& 13 	& 144.68 \\
    Reachability 		& 0  	& US 		& 0	& US 		& 20 	& 145.86 	& 6  	& 140.56 \\
    Ricochet Robots 		& 7  	& 60.76 	& 6	& 87.46 	& 9 	& 71.48 	& 7  	& 94.92 \\
    Sokoban 			& 10 	& 90.08 	& 9	& 145.24 	& 8  	& 78.51 	& 8  	& 92.85 \\
    Stable Marriage 		& 4  	& 402.30 	& 7	& 363.72 	& 5  	& 387.66 	& 2  	& 402.24 \\
    Steiner Tree 		& 2  	& 52.12 	& 1	& 270.88 	& 2  	& 40.98 	& 1  	& 132.88 \\
    Strategic Companies   	& 0  	& US 		& 0	& US 		& 18 	& 151.30 	& 7  	& 30.30 \\
    System Synthesis 		& 0  	& TO 		& 0  	& TO		& 0  	& TO 		& 0  	& TO \\
    Valves Location Problem 	& 16 	& 42.91 	& 15	& 58.69 	& 16 	& 42.70 	& 15 	& 40.01 \\
    Video Streaming 		& 13 	& 57.93 	& 0	& TO 		& 13 	& 62.61 	& 9  	& 8.99 \\
    Visit-all 			& 8  	& 17.26 	& 8	& 62.91 	& 8 	& 14.93 	& 8  	& 65.43 \\
\hline
    \textbf{Total Solved Instances} 	& 254 	&  	& 233 	& 	& 330 	& 	& 263 \\
\hline\hline
    \end{tabular}
\end{table}

\medskip
Results evidence, as already noted by looking at the plot of Figure~\ref{fig:solving}, good performance of the two solvers when combined with \idlv: indeed, \clasp and \wasp solve a total of $330$ and $263$ instances, respectively, instead of a total of $254$ and $233$, respectively, when combined with \gringo.
It is worth noting that the noticeable difference is due also to the explicit support for queries, which is absent in \gringo; however, performance of the combinations based on \idlv remain satisfactory even if domains with queries are not taken into account: $272$ and $232$ instances for \clasp and \wasp, respectively.
This is encouraging, and suggests that \idlv can be used as a solid base for a competitive ASP system.

The positive impact of \idlv on the two solvers
can be explained by the many optimizations that \idlv employs for manipulating input rules in order to decrease their intrinsic complexity, that often lead to the production of a smaller number of rules that feature smaller body lengths, with respect to what produced by \gringo.
This explains, for instance, the performance gain enjoyed by the solvers in the case of {\em Permutation Pattern Matching}, due to a careful automatic optimization of the input programs performed by \idlv by means of the heuristic-guided decomposition rewriting.
In other situations, the difference is made by the impact of the grounding on the solving heuristics adopted during the answer set search, such as in the case of {\em Video Streaming} for \wasp, which benefits from a different rule order produced by \idlv.

However, a closer look to performance of the solvers and their different behaviors, allows us to make some additional considerations.
We first note that \clasp outperforms \wasp on $10$ benchmarks: {\em Abstract Dialectical Frameworks}, {\em Combined Configuration}, {\em Complex Optimization}, {\em Graph Coloring}, {\em Incremental Scheduling}, {\em Partner Units}, {\em Reachability}, {\em Ricochet Robots}, {\em Strategic Companies}, and {\em Video Streaming}; on the other hand, \wasp outperforms \clasp on {\em Connected Still Life}, {\em Crossing Minimization}, {\em Maximal Clique}, and {\em MaxSAT}.
In general, \clasp results to be more efficient than \wasp in problems involving cyclic disjunctive rules, such as {\em Abstract Dialectical Frameworks}, {\em Complex Optimization} and {\em Strategic Companies}; this difference is mainly due to the inefficient implementation of unfounded-set check performed by \wasp, which is required for such problems.
Another important difference between the two solvers is the default algorithm employed on optimization problems.
The default configuration of \clasp used in our experiments implements a model-guided algorithm, whose idea is to search for any answer set to initialize an overestimation of the optimal cost, and then to improve the overestimation by searching new answer sets of improved cost, until the overestimation cannot further improved.
The default version of \wasp, instead, is based on a core-guided algorithm:
the idea is to start by searching an answer set satisfying all weak constraints, which would be therefore optimal; if such an answer set does not exist, a subset of the weak constraints that cannot be jointly satisfied, called \textit{unsatisfiable core}, is identified and subsequently relaxed so to search for answer sets satisfying all but one of the original weak constraints.
The core-guided algorithm of \wasp is complemented with the \textit{stratification}~\cite{doi:10.1093/logcom/exv061} technique which first considers only weak constraints of greatest weight.
After all cores have been analyzed, a (suboptimal) answer set will be found, and weak constraints with the second greatest weight will be also considered.
As already noted in~\cite{doi:10.1093/logcom/exv061}, model-guided and core-guided algorithm show complementary behavior, and for such reason \clasp shows better performance on {\em Video Streaming}, while the core-guided algorithm implemented by \wasp obtains better results on instances of {\em Connected Still Life}, {\em Crossing Minimization}, {\em Maximal Clique and MaxSAT}.

Therefore, none of the solvers is the best on all domains; this proves that there is room for improvements by means of proper smart selection mechanisms.
Indeed, if we focus on \idlv combined with the two different solvers, it is possible to observe that there are $76$ and $30$ instances uniquely solved by \clasp and \wasp, respectively.

\section{Automatic coupling \idlv with a solver }\label{sec:selector}
In this section, we illustrate how \idlvs has been designed. The aim is obtaining a new ASP system that combines \idlv with an automatic solver selector, that is able to automatically classify instances as ``better suited for \clasp'' or ``better suited for \wasp'' on the basis of some inherent features of their instantiations as produced by \idlv.
In other words, the goal is to perform such identification thanks to a learning phase relying on  a \emph{training set} consisting of a pair of observations on some features and  membership to a category.
Formally, the observations (or instances) consist of an input vector $X = \{x_1, \dots, x_n \} $ with $x_i \in \mathbb{R}^k$, $i\in\{1 \dots n\}$, being a vector of values over $k$ features; the categories (or labels) consist of a vector $Y = \{y_1, \dots, y_m\}$, where $m$ is the number of the classes of the classification problem; in our scenario, $m$ is the number of the ASP solvers.
Hence, our training set consists of a pair $(X,Y)$, where to each instance $x_i\in X$ is associated the ``best'' solver $y_j\in Y$.
The \emph{training phase} is the task of generating a classifier implementing $c : \mathbb{R}^k \rightarrow Y$ that to each  $x_i \in X$ properly associates $y_j\in Y$.
 In order to build the classifier, we considered four different approaches that have been compared in order to choose the best suited to our scenario.


\subsection{Dataset}\label{subsec:dataset}

In order to build the dataset, we started by registering the performance of the solvers over all benchmarks submitted to the $6th$ ASP Competition~\cite{DBLP:journals/jair/GebserMR17} as grounded by \idlv.
As usual in such cases, the dataset has been divided into: \emph{training}, \emph{validation} and \emph{test} sets,
distributed with the ratio of 50\%, 25\%, and 25\%, respectively, in order to tackle overfitting.
Moreover, since the label distribution is unbalanced (\clasp  solves significantly more instances than \wasp), we used a {\em stratified sampling}\cit{DBLP:journals/bioinformatics/EsfahaniD14} in order to ensure that relative labels frequencies are approximately preserved in each split.

\subsection{Features Selection}
In the following we describe the features that have been considered for the classification.

The idea was to identify a set of features which could somehow ``describe'' the ground program at hand, and that, in the meanwhile, would be independent from any solver implementation detail, so that it could be eventually easy to test the machinery with additional systems.
We discarded all features that are not computable in linear time, as we wanted to keep the overhead as little as possible: we recall here that we compute them over the ground program, whose size can be considerably large, in the general case.
Then, we decided to find proper descriptions for two fundamentals aspects of ASP programs, namely \emph{atoms ratio} and \emph{rules ratio}; among several possibilities, according to the result of a thorough experimental analysis, we identified the following features as the most suitable:
\begin{itemize}
  \item
    \emph{Atoms ratios.}\ We considered five different ratios that represent the type of atoms and a raw measure of their distribution in the input ground program:
  $$ (a): \frac{F}{R} \ \ \ \ \  (b): \frac{PA}{R} \ \ \ \ \ (c): \frac{NA}{R} \ \ \ \ \ (d): \frac{PA}{BA} \ \ \ \ \ (e): \frac{NA}{BA} $$
    where $F$ is the total number of facts and always true atoms, $R$ the total number of ground rules, $PA/NA$ the total number of positive/negative atoms and $BA$ the total number of atoms appearing in rule bodies.
  \item
    \emph{Rules ratios.}\ We considered five different ratios that represent the type of rules and a raw measure of their distribution in the input ground program, taking into account also advanced constructs of the ASP-Core-2 standard language~\cite{calimeri2012asp}, such as choices, aggregates, and weak constraints:
  $$ (f): \frac{C}{R} \ \ \ \ \  (g): \frac{W}{R} \ \ \ \ \ (h): \frac{SR}{R} \ \ \ \ \ (i): \frac{CR}{R} \ \ \ \ \ (j): \frac{WR}{R} $$
    where $C$ is the total number of strong constraints, $W$ the total number of weak constraints, $SR$ the total number of standard rules, $CR$ the total number of choice rules and $WR$ the total number of weight rules.
    Please note that with standard rules we denote rules without aggregate or choice atoms; weight and choice rules, instead, handle aggregate literals and choice atoms generated by the grounder.
\end{itemize}

\subsection{Choice of a Classification Algorithm}

The considered classifiers are the following:

\begin{itemize}
  \item {\em Random forest}: combination of tree classifiers such that each tree depends on the values of a random input vector sampled independently and with the same distribution for all trees in the forest. This classifier uses averaging to improve the predictive accuracy and control over-fitting\cit{DBLP:journals/ml/Breiman01}.

  \item {\em Neural Network}: a multi-layer perceptron (MLP) algorithm that trains using back\hyp{}propagation. The classifier can learn a non-linear function approximator for classification problem; between the input and the output layer, there can be one or more non-linear layers, called hidden layers\cit{DBLP:journals/ml/Breiman01}.

  \item {\em Support vector machine} (SVM):  a binary linear classifier that basically tries to find an optimal linear hyperplane such that the expected classification error is minimized\cit{smola2004tutorial}.

  \item {\em AutoWEKA}: automatic classifier selection and hyperparameter optimization that try to identify the best classifier algorithm and appropriate hyperparameter settings\cit{DBLP:journals/jmlr/KotthoffTHHL17}.
\end{itemize}

After the training and the hyperparameter tuning phases, we used the \emph{precision-recall metric} to evaluate output quality of the classifiers: precision is a measure of how many relevant results appear within those that are returned, while recall is a measure of how many relevant results are actually returned (i.e., not missed).
In particular, we computed the traditional $F_1$ score as the harmonic mean of precision and recall:
$$\frac{ 2 * Precision * Recall}{ Precision + Recall}$$
in Table\re{clas-res} shows the precision, recall and $F_1$ score for each classifiers on the test set.
The overall results show good precision and recall for all candidates; however, \emph{AutoWEKA} and \emph{Random Forest} feature lower precision when compared to the \emph{Neural Network} and the \emph{SVM}; even if the latter two perform similarly, we decided to choose the \emph{SVM} as in our tests it required less time to predict new instances.

\begin{table}[b!]
\centering
\caption{Results of the classifiers on test set.}
\label{clas-res}
\begin{tabular}{lrrr}
\hline\hline
              & \textbf{Precision} & \textbf{Recall} & \textbf{$F_1$ score}  \\
\hline
Neural Network     & 0.94               & 0.87            & 0.89               \\
Random Forest & 0.89               & 0.87            & 0.88               \\
SVM           & 0.94               & 0.87            & 0.89               \\
AutoWEKA      & 0.88               & 0.87            & 0.87              \\
\hline\hline
\end{tabular}
\end{table}

\subsection{The \idlvs System}\label{subsec:idlvs}
\begin{figure}[h!]
 \begin{center}
  \includegraphics[width=0.98\textwidth]{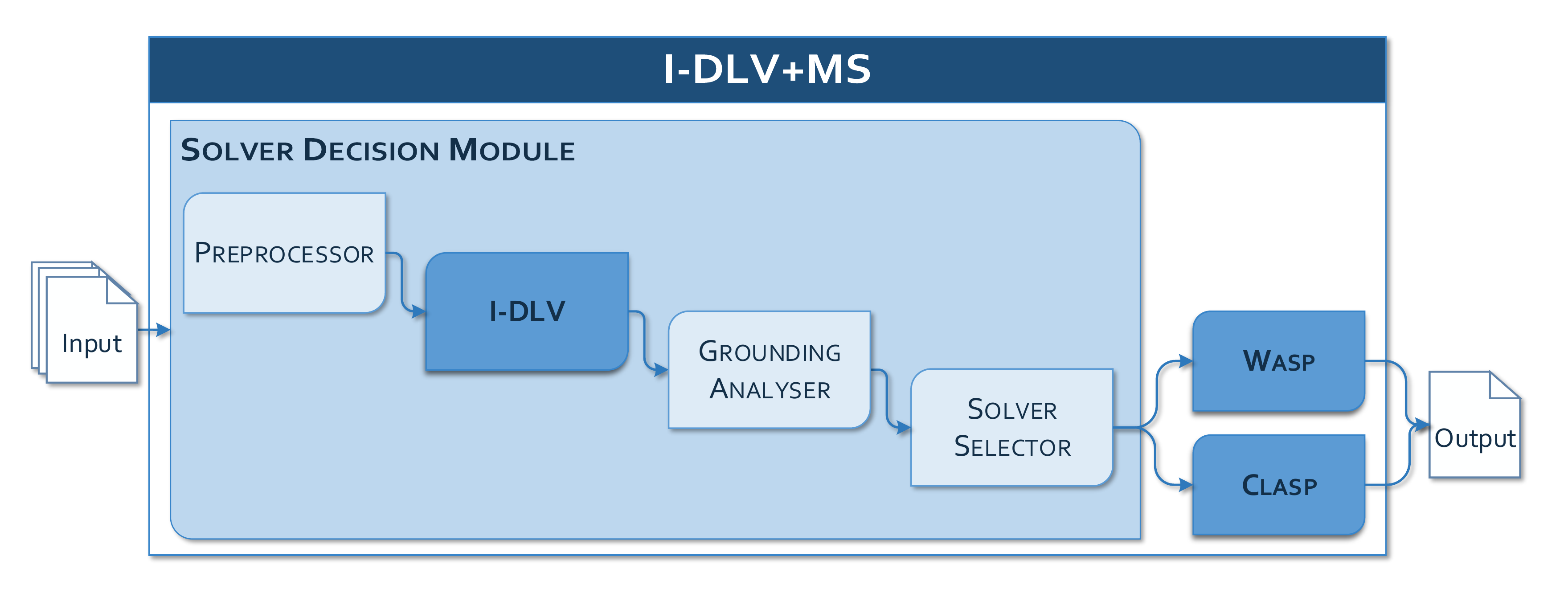}
  \caption{{\small \idlvs Architecture.}}
  \label{fig:Arch}
 \end{center}
\end{figure}

The architecture of \idlvs is reported in Figure~\ref{fig:Arch}.
The \preproc module analyzes the input program \emph{P},
and interacts with the \idlv system in order to determine if the input program is non-disjunctive and stratified, as these kinds of programs are completely evaluated by \idlv without the need for a solver.
If this is not the case, the \stats module extracts the intended features from the ground program produced by \idlv and passes them to the classification module.
Then, the \selector tries to foresee, among the available solvers, which one would perform better, and selects it.

Even if the current version of \idlvs supports just \clasp and \wasp, its modular architecture easily allows one to update the solvers or even add additional ones.
Clearly, such changes would require the prediction model to be retrained with appropriate statistics on the new solvers.
In particular, it is worth mentioning that, since \clasp and \wasp offer several command-line options that modify some aspects of their computation, in order to find the best way of coupling them with \idlv, we first performed an experimental activity
by testing the two solvers with different combinations of such command-line options, before the selector was tuned.
For the sake of readability, we omit to extensively report here the results obtained in this preliminary step; we just point out that a proper analysis of the results suggested to select the option \texttt{--configuration=trendy} for \clasp, which basically changes the heuristic parameters used during the search, and options \texttt{--shrinking\hyp strategy=progression} \texttt{--shrinking\hyp budget=10 --trim-core --enable\hyp disjcores} for \wasp.
It is worth noting that such options force \wasp to use two techniques tailored to optimization problems, namely \textit{disjoint cores preprocessing} and \textit{core-shrinking}~\cite{DBLP:journals/tplp/AlvianoD16}, and that are not supported by \clasp; even if core-shrinking is now supported also by \clasp since version $3.3.0$, it was not available at the submission time of the $7th$ ASP competition, when \idlvs was being designed and implemented.

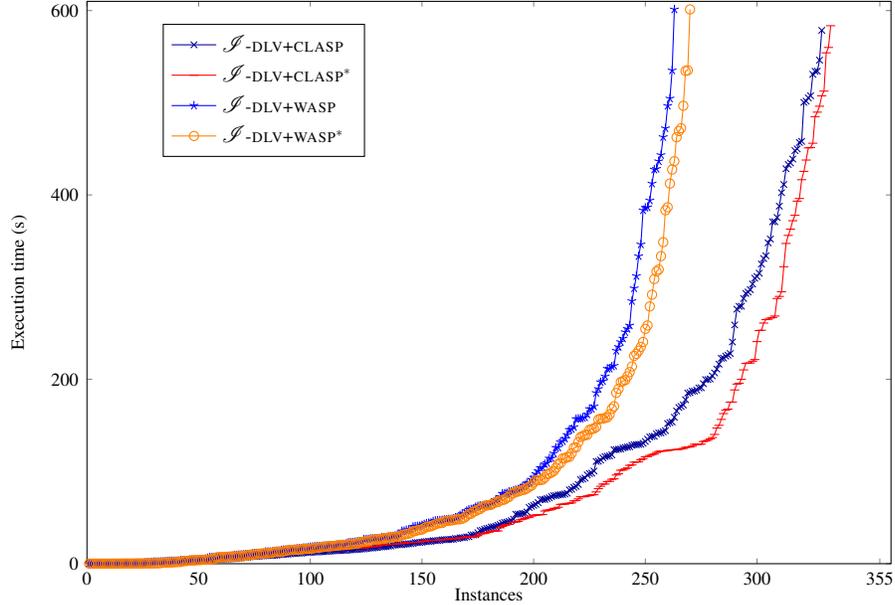
\begin{figure}[t!]\centering
	\begin{tikzpicture}[scale=0.99]
	\pgfkeys{%
		/pgf/number format/set thousands separator = {}}
	\begin{axis}[
	scale only axis
	, font=\scriptsize
	, x label style = {at={(axis description cs:0.5,0.04)}}
	, y label style = {at={(axis description cs:0.03,0.5)}}
	, xlabel={Instances}
	, ylabel={Execution time (s)}
	, width=0.8\textwidth
    , legend style={at={(0.22,0.96)},anchor=north}
	, height=0.56\textwidth
	, ymin=0, ymax=610
	, xmin=0, xmax=360
	, xtick={0,50,100,150,200,250,300,355}
	, ytick={0,200,400,600}
	, major tick length=2pt
	]
	\addplot [mark size=1.75pt, color=blue!60!black, mark=x] [unbounded coords=jump] table[col sep=semicolon, y index=1] {./idlv-solvers-options.csv};
    \addlegendentry{\idlva+\clasp}
	\addplot [mark size=1.75pt, color=red, mark=-] [unbounded coords=jump] table[col sep=semicolon, y index=2] {./idlv-solvers-options.csv};
	\addlegendentry{\idlva+\claspmod}
	\addplot [mark size=1.75pt, color=blue, mark=star] [unbounded coords=jump] table[col sep=semicolon, y index=3] {./idlv-solvers-options.csv};
	\addlegendentry{\idlva+\wasp}	
	\addplot [mark size=1.75pt, color=orange, mark=o] [unbounded coords=jump] table[col sep=semicolon, y index=4] {./idlv-solvers-options.csv};
    \addlegendentry{\idlva+\waspmod}	
\end{axis}
	\end{tikzpicture}
\caption{Experiments for the empirical selection of invocation options for \clasp and \wasp when coupled with \idlv, performed on benchmarks from the $6th$ ASP Competition.\label{fig:opzioni-solvers}}
\end{figure}

The cactus plot in Figure~\ref{fig:opzioni-solvers} evidences the benefits coming from the usage of the discussed options instead of the default configurations: it compares  $(i)$\idlva+\clasp (\clasp invoked with the default configuration), $(ii)$ \idlva+\claspmod\ (\clasp invoked with the selected configuration), $(iii)$\idlva+\wasp (\wasp invoked with the default configuration), and $(iv)$ \idlva+\waspmod\ (\wasp invoked with the selected configuration); the number of solved instances is on the $x$-axis while running times (in seconds) are on the $y$-axis.
Consequently, the selector has been actually trained on the basis of the performance of \claspmod\ and \waspmod, which, in turn have been incorporated into the system as depicted in Figure~\ref{fig:Arch}. 

\idlvs is freely available at~\cite{idlv-ms-web}.

\section{Experimental Evaluation of \idlvs}\label{sec:experiments}
In this section, we report the results of an experimental analysis conducted in order to empirically assess performance of \idlvs.
\begin{table}[t]
  \centering
    \caption{Comparison of \idlvs against \idlva+\claspmod\ and \idlva+\waspmod\ on benchmarks from the $6th$ ASP Competition (20 instances per problem): number of solved instances and average running times (in seconds).
    Time outs and unsupported syntax issues are denoted by TO and US, respectively.
    \label{tab:idlvs-options}}
    \tabcolsep=0.050cm
    \begin{tabular}{lrrrrrr}
\hline\hline
    \multirow{2}[2]{*}{\textbf{Problem}} & \multicolumn{2}{c}{\textbf{\idlva+\claspmod}} & \multicolumn{2}{c}{\textbf{\idlva+\waspmod}} & \multicolumn{2}{c}{\textbf{\idlvs}} \\
       & \textbf{\#solved} & \textbf{time} & \textbf{\#solved} & \textbf{time} & \textbf{\#solved} & \textbf{time} \\
    \cmidrule{2-3}     \cmidrule{4-5}      \cmidrule{6-7}
    \hline
    Abstract Dial. Frameworks & 20 & 8.46 & 14 & 27.01 & 20 & 9.61 \\
    Combined Configuration & 4  & 46.60 & 0  & TO & 4  & 51.14 \\
    Complex Optimization & 19 & 115.11 & 6  & 152.15 & 19 & 118.28 \\
    Connected Still Life & 5  & 290.71 & 10 & 98.31 & 10 & 87.05 \\
    Consistent Query Ans. & 20 & 86.37 & 18 & 87.34 & 20 & 102.42 \\
    Crossing Minimization & 7  & 71.20 & 19 & 1.02 & 19 & 2.30 \\
    Graceful Graphs & 9  & 125.49 & 6  & 194.57 & 8  & 68.63 \\
    Graph Coloring & 16 & 102.89 & 8  & 133.42 & 16 & 103.66 \\
    Incremental Scheduling & 11 & 36.75 & 6  & 77.94 & 11 & 41.68 \\
    Knight Tour With Holes & 14 & 26.37 & 10 & 34.85 & 14 & 28.43 \\
    Labyrinth & 10 & 125.90 & 10 & 70.98 & 10 & 128.60 \\
    Maximal Clique & 0  & TO & 15 & 152.43 & 15 & 157.62 \\
    MaxSAT & 9  & 97.56 & 19 & 98.78 & 18 & 99.34 \\
    Minimal Diagnosis & 20 & 14.81 & 20 & 24.45 & 20 & 17.65 \\
    Nomistery & 8  & 25.38 & 8  & 36.73 & 8  & 26.66 \\
    Partner Units & 14 & 25.24 & 5  & 131.61 & 14 & 26.36 \\
    Permutation Pattern Matching & 20 & 19.20 & 20 & 22.31 & 20 & 24.82 \\
    Qual. Spatial Reasoning & 20 & 108.87 & 13 & 144.68 & 20 & 115.41 \\
    Reachability & 20 & 145.86 & 6  & 140.56 & 20 & 155.86 \\
    Ricochet Robots & 10 & 101.86 & 7  & 94.92 & 11 & 144.63 \\
    Sokoban & 9  & 78.53 & 8  & 92.85 & 8  & 91.68 \\
    Stable Marriage & 9  & 333.31 & 2  & 402.24 & 8  & 418.16 \\
    Steiner Tree & 3  & 179.96 & 0  & TO & 3  & 180.19 \\
    Strategic Companies   & 18 & 151.30 & 7  & 30.30 & 18 & 154.29 \\
    System Synthesis & 0  & TO & 0  & TO & 0  & TO \\
    Valves Location Problem & 16 & 12.45 & 16 & 69.44 & 16 & 14.95 \\
    Video Streaming & 14 & 64.40 & 9  & 0.11 & 14 & 65.05 \\
    Visit-all & 8  & 16.78 & 8  & 65.43 & 9  & 81.88 \\
\hline
    \textbf{Total Solved Instances} & \textbf{333} &    & \textbf{270} &    & \textbf{373} &  \\
\hline\hline
    \end{tabular}%
\end{table}

We first compare \idlvs against \idlv combined with the two solvers configured as described in Section~\ref{subsec:idlvs}, i.e., \claspmod\ and \waspmod.
The results of such comparison are reported in
Table\re{tab:idlvs-options}.
\idlvs performs better than both \idlva+\claspmod\ and \idlva+\waspmod; in particular, it solves $40$ and $103$ instances more than \idlva+\claspmod\ and \idlva+\waspmod, respectively.
More in detail, when comparing \idlvs against \idlva+\claspmod, even if \idlvs ``misses'' a solvable instance in {\em Graceful Graphs}, {\em Sokoban} and {\em Stable Marriage} with respect to \idlva+\claspmod, a noticeable advantage emerges in several other domains, such as {\em Connected Still Life}, {\em Crossing Minimization}, {\em Maximal Clique}, {\em MaxSAT}, {\em Ricochet Robots}, {\em Visit-all}, where \idlvs solves $5$, $12$, $15$, $9$, $1$, $1$ additional instances, respectively.
When comparing \idlvs against \idlva+\waspmod, the improvements achieved by \idlvs are even more evident, with peaks of $13$ instances for {\em Complex Optimization}, $9$ instances for {\em Partner Units}, $14$ instances for {\em Reachability} and $11$ instances for {\em Strategic Companies}; the only exception is {\em MaxSAT} in which \idlva+\waspmod\ solves one instance more than \idlvs.

A more careful analysis of Table~\ref{tab:idlvs-options} might lead to the conclusion that the number of instances solved by \idlvs, for each domain, is limited by the best configuration \idlva+\claspmod\ and \idlva+\waspmod, i.e., as if it chooses the same solver for all instances of a given domain.
This is actually the case for some benchmarks, as, for instance, \emph{Combined Configuration}, where it is evident that the selector always elects \claspmod.
This is not surprising, as most of the ASP competition instances of a given benchmark problem, when coupled with the encoding, produces ground programs that feature similar structure; hence, same values for the metrics employed by the selector are induced.
Nevertheless, this is not true in general, as witnessed by cases such as \emph{Ricochet Robots}, \emph{Graceful Graphs}, and others.

The smart selection machinery causes an overhead; this is expected, as it requires to analyze the produced instantiation in order to first compute the features needed for selecting the solver and then properly invoking it.
However, such overhead is definitely acceptable, as it can be clearly observed by taking into account domains for which only one among \idlva+\claspmod\ and \idlva+\waspmod\ is able to solve at least one instance.
For example, when dealing with {\em Combined Configuration} and {\em Steiner Tree}, \idlvs performs analogously to \idlva+\claspmod, while when dealing with {\em Maximal Clique} it reflects the behaviour of \idlva+\waspmod, solving the same number of instances with an overhead that goes from less than $1\%$ to no more than $9\%$.

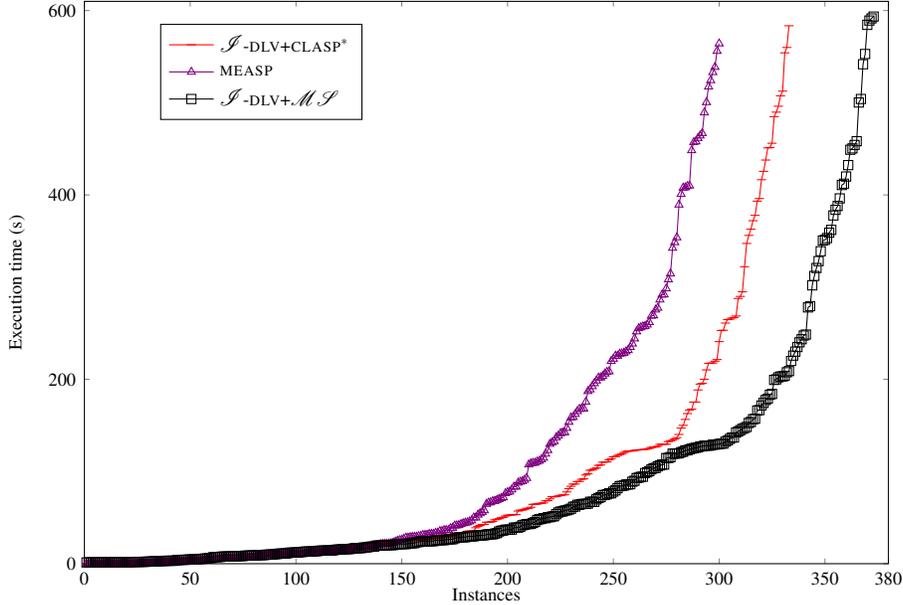
\begin{figure}[t!]\centering
	\begin{tikzpicture}[scale=0.99]
	\pgfkeys{%
		/pgf/number format/set thousands separator = {}}
	\begin{axis}[
	scale only axis
	, font=\scriptsize
	, x label style = {at={(axis description cs:0.5,0.04)}}
	, y label style = {at={(axis description cs:0.03,0.5)}}
	, xlabel={Instances}
	, ylabel={Execution time (s)}
	, width=0.8\textwidth
    , legend style={at={(0.22,0.96)},anchor=north}
	, height=0.56\textwidth
	, ymin=0, ymax=610
	, xmin=0, xmax=380
	, xtick={0,50,100,150,200,250,300,350,380}
	, ytick={0,200,400,600}
	, major tick length=2pt
	]
	\addplot [mark size=1.75pt, color=red, mark=-] [unbounded coords=jump] table[col sep=semicolon, y index=1] {./ASP-2015-8.csv};
    \addlegendentry{\idlva+\claspmod}
	\addplot [mark size=1.75pt, color=violet, mark=triangle] [unbounded coords=jump] table[col sep=semicolon, y index=2] {./ASP-2015-8.csv};
	\addlegendentry{\measp}	
	\addplot [mark size=1.75pt, color=black, mark=square] [unbounded coords=jump] table[col sep=semicolon, y index=3] {./ASP-2015-8.csv};
    \addlegendentry{\idlvs}	
\end{axis}
	\end{tikzpicture}
\caption{Experimental evaluation of \idlvs on benchmarks from the $6th$ ASP Competition.\label{fig:asptest}}
\end{figure}

Further experiments have been performed for testing \idlvs against $(i)$ \idlva+\claspmod, which is the best combination of \idlv with a solver obtainable without the use of a smart selection machinery, and $(ii)$ the latest available version of \measp\footnote{\url{http://aspcomp2015.dibris.unige.it/participants}}, which represents the state-of-the-art among multi-engine ASP solvers, and was also the winner of the $6th$ ASP Competition.
Results of such comparison are depicted in the cactus plot reported in Figure~\ref{fig:asptest}, where the number of solved instances is on the $x$-axis while running times (in seconds) are on the $y$-axis.
In general, \idlvs outperforms both \idlva+\claspmod and \measp.
While, according to the consideration above, the results of the comparison against \idlva+\claspmod\ are expected, some additional considerations deserve to be made about \measp.

Differently from \idlvs, \measp employs machine-learning techniques for the inductive choice of the best solver on a per instance basis.
Moreover, even if there are similarities with \idlvs, there are some other key differences, starting from the nature and the number of ASP systems employed.
Indeed, in order to choose the solver, \measp relies on the instantiation produced by \gringo; furthermore, it chooses among five solvers, way more than the mere two taken into account by \idlvs.
On the one hand, given that different engines use evaluation strategies that can be substantially different, the use of a larger pool of solvers allows to gain a significant number of instances that are uniquely solvable by one solver.
On the other hand, such differences imply that a high price might be paid in case of a wrong choice; in fact, if the space for choices is narrowed, the probability of making a wrong one decreases, and this might lead to a more consistent behavior, as in the case of \idlvs.

In addition, it might be interesting to note that \idlva+\claspmod\ outperforms \measp.
This can be partially explained by observing that, as already mentioned, \measp relies on \gringo, for grounding, and the best configuration overall obtainable from its pool of solvers consist of \gringo and \clasp; however, we already noted in Section~\ref{sec:idlv} how \clasp performance look to increase when working on \idlv instantiations instead of on \gringo ones.

A final consideration concerns overfitting, which can be an important issue when dealing with inductive machineries that rely on training.
As pointed out in Section~\ref{subsec:dataset}, we specifically designed the dataset in order to prevent the problem, and the system appears to be able to adapt to unseen situations: indeed, the selector was trained on the basis of the data available from the $6th$ ASP Competition, and participated in the $7th$, which featured significant differences with respect to the previous one in terms of new domains, new encodings for already present domains and new instances.
Nevertheless, \idlvs was able to perform well, and even win the regular track.

%
%


\section{Related Work}\label{sec:related}
Traditional evaluation strategy of ASP systems is based on two stages and mimics the intended semantics by first performing \textit{grounding} and then executing \textit{solving}; for both phases, several efficient systems have been proposed during the years.

As for instantiation, the most widespread grounders are (the one in) \dlv~\cite{DBLP:conf/birthday/FaberLP12} and \gringo~\cite{DBLP:conf/lpnmr/GebserKKS11}; both systems are based on semina\"ive database evaluation techniques~\cite{DBLP:books/cs/Ullman88} for avoiding duplicate work during grounding, and accept safe programs as input~\cite{DBLP:conf/birthday/FaberLP12,DBLP:journals/ia/CalimeriFPZ17,DBLP:conf/lpnmr/GebserKKS11}.
Other approaches have been implemented in \lparse~\cite{DBLP:conf/lpnmr/Syrjanen01} and in earlier versions of
\gringo~\cite{DBLP:conf/lpnmr/GebserST07}, that take as input $\omega$- or $\lambda$-restricted programs, respectively.
\idlv~\cite{DBLP:journals/ia/CalimeriFPZ17} has been released as the new version of the \dlv grounder; it shares with \dlv and \gringo the support to safe programs and the general evaluation schema based on semina\"ive; however, it significantly differs in the optimization strategies from both of them and, furthermore, differently from \dlv, it fully supports the ASP-Core-2~\cite{calimeri2012asp} language standard.

Concerning ASP solvers, the first generation, i.e. \textsc{smodels}~\cite{DBLP:journals/ai/SimonsNS02} and \dlv \cite{DBLP:journals/tocl/LeonePFEGPS06,DBLP:journals/jal/MarateaRFL08}, was based on a DPLL-like algorithm extended with inference rules specific to ASP.
Modern ASP solvers, including \clasp~\cite{DBLP:conf/iclp/GebserKKOSW16,DBLP:journals/ai/GebserKS12,DBLP:conf/lpnmr/GebserKK0S15} and \wasp~\cite{DBLP:conf/cilc/DodaroAFLRS11,DBLP:conf/lpnmr/AlvianoDLR15}, also include mechanisms for conflict-driven clause learning and for non-chronological backtracking.

\medskip

Interestingly, the attempt to find ways of enjoying good performance over different problem domains by means of smart choices has already been pursued by neighbouring communities, by means of proper strategies of \emph{algorithm selection}~\cite{DBLP:journals/ac/Rice76}; we cite here, for instance, what has been done for solving propositional satisfiability (SAT)~\cite{DBLP:journals/jair/XuHHL08} and Quantified SAT (QSAT)~\cite{DBLP:journals/constraints/PulinaT09}.
%
As for what ASP is concerned, some interesting works in this respect have already been carried out in~\cite{DBLP:journals/tplp/MarateaPR14} and~\cite{DBLP:journals/tplp/HoosLS14}: the former proposal is the well-established multi-engine ASP system \measp, the winner of the 6th ASP Competition, while the latter  
relies on a single solver, i.e. \clasp, and uses similar strategies in order to select the ``best'' configuration for it.
Some considerations about similarities and differences between such proposals and \idlvs follow.

\medskip

\idlvs is a portfolio solver and its working principles are similar to the ones of the multi engine ASP solver \measp~\cite{DBLP:journals/tplp/MarateaPR14}. Indeed, both solvers are based on a classification of instances according to their internal features.
In particular, the ones computed by \measp can be classified in four different categories:
\textit{(i) problem size}, i.e. number of rules $r$, number of atoms $a$, ratios $(r/a)$, $(r/a)^2$, $(r/a)^3$, $(a/r)$, $(a/r)^2$, $(a/r)^3$;
\textit{(ii) balance}, i.e. fraction of unary, binary, and ternary rules;
\textit{(iii) proximity to Horn}, i.e. fraction of Horn rules and number of occurrences in a Horn rule for each atom;
\textit{(iv) ASP specific}, i.e. number of rules with empty bodies, fraction of normal rules and constraints.
As a design choice, all the features considered by \measp can be easily computed, that is, they are computable in linear time in the size of the input program.
\idlvs considers a different set of features, as detailed in Section~\ref{sec:selector}, still linear-time computable; they also include rule types, e.g., choice and weight rules.
Additional differences concern the pool of engines employed by \measp and \idlvs.
As already mentioned in Section~\ref{sec:experiments}, the former uses \clasp(and its variant \textsc{claspd}; \citeN{DBLP:conf/lpnmr/GebserKK0S15}), \cmodels~\cite{DBLP:conf/lpnmr/LierlerM04}, \dlv~\cite{DBLP:journals/tocl/LeonePFEGPS06}, and \textsc{idp}~\cite{DBLP:journals/tplp/BruynoogheB0CPJ15}, while \idlvs uses only \clasp~\cite{DBLP:conf/lpnmr/GebserKK0S15} and \wasp~\cite{DBLP:conf/lpnmr/AlvianoDLR15}.

\idlvs is also related to \textsc{claspfolio}~\cite{DBLP:journals/tplp/HoosLS14}, which selects the \textit{best} configuration for the solver \clasp.
The first difference between the two solvers is related to the inductive model; indeed, the one of \textsc{claspfolio} is based on \textit{regression} techniques, while the inductive model of \idlvs is based on \textit{classification} algorithms.
Moreover, \textsc{claspfolio} uses dynamic features that are computed at each restart, e.g. the number of choices performed by the solver, while \idlvs only considers static features.
Finally, since \textsc{claspfolio} is tailored to \clasp, it takes into account features that cannot be computed by \wasp, e.g. the number of equivalences.
On the contrary, \idlvs considers the underlying solvers as black boxes, therefore it can be extended for including also other solvers other than \clasp and \wasp.

\section{Conclusions and Future Work}\label{sec:conclusions-ongoing}

In this work we presented \idlvs, a new ASP system that relies on a modern ASP grounder, namely \idlv, and of a smart selection machinery to inductively choose a solver depending on the features of the input at hand.
We reported the rationales and the details about design and implementation of the system, along with a number of experimental analyses.
The system shows good performance, both when compared against \measp, which was the winner of the $6th$ ASP Competition and makes use of similar inductive approaches for selecting ASP solvers, and when compared against combinations of \idlv with the mainstream ASP solvers \clasp and \wasp.
Good performance is confirmed also by the success in the latest ($7th$) ASP competition~\cite{DBLP:conf/lpnmr/GebserMR17}, where the system won the regular track in the $SP$ category.

\medskip


Note that all systems employed and tested in this work refer to the respective versions available at the time of designing, implementing and training the herein proposed system, i.e., before the $7th$ ASP Competition officially started.
At the time of finalizing this manuscript, new versions came out for many of such systems, both solvers and grounders.
As future work, we plan to update \idlvs by taking into account the new versions, as results and considerations might significantly change.
Furthermore, we will take into account additional ASP solvers with different parameterizations, and explore more features for improving classification capabilities and achieve better overall performance.
Eventually, we are studying the possibility of designing a built-in automatic configurator for \idlv that, using machine-learning techniques, properly sets a bunch of optimization parameters among the several currently available in \idlv; this would make the grounder capable to automatically adapt to the problem at hand.


\bibliography{references-cleaned}
\end{document}